\documentstyle[11pt,newpasp,twoside,epsf]{article}
\def\edcomment#1{\iffalse\marginpar{\raggedright\sl#1\/}\else\relax\fi}
\reversemarginpar

\begin{document}
\title{Spectrally resolved flares in the quiescent black hole V404~Cyg} 

\author{R. I. Hynes, P. A. Charles} 

\affil{Department of Physics and Astronomy, University of
Southampton, Southampton, SO17 1BJ, UK; rih@astro.soton.ac.uk}

\author{C. Zurita, J. Casares} 

\affil{Instituto de Astrof\'\i{}sica de Canarias, 38200 La Laguna, 
Tenerife, Spain}

\author{C. A. Haswell, D. A. Lott} 

\affil{Department of Physics and Astronomy, The Open University, Walton
Hall, Milton Keynes, MK7 6AA}

\begin{abstract}
We present a spectrophotometric study of short-term optical
variability in the quiescent black hole X-ray transient V404~Cyg
focusing on two nights of H$\alpha$ spectroscopy.  We find
significant variability, with both the H$\alpha$ line and the
continuum varying in a correlated way.  This includes both dramatic
flares lasting a few hours in which the line flux nearly doubles and
lower-level flickering.  The strongest flares involve development of
asymmetry in the line profile, with the red wing usually strongest
independent of orbital phase.  Based on the line profile changes
during the flares, we conclude that the most likely origin for the
variability is variable photoionisation by the central source.
\end{abstract}

The quiescent black hole binary V404~Cyg has long been known to
exhibit significant variability in quiescent optical, IR, X-ray and
radio fluxes (see Hynes et al.\ (2002) for summary and references).
In particular, H$\alpha$ line profiles change significantly on short
timescales (Casares \& Charles 1992; Casares et al.\ 1993).
Understanding these variations is important as the properties of the
variability can yield clues about the nature of the quiescent
accretion flow and provide an observational test of models for this
flow.  To further investigate this behaviour, in 1999 July we obtained
red spectroscopy at high time-resolution so as to reveal the spectral
signature of the short-term variability.  We obtained simultaneous
photometry to ensure a reliable flux calibration; hence we can
accurately study both line and continuum variations.  The low spectral
resolution also provides us with kinematic resolution of emission line
variations.  A thorough analysis of these data is given in Hynes et
al.\ (2002); here we summarise some of the key results.

\begin{figure}
\plottwo{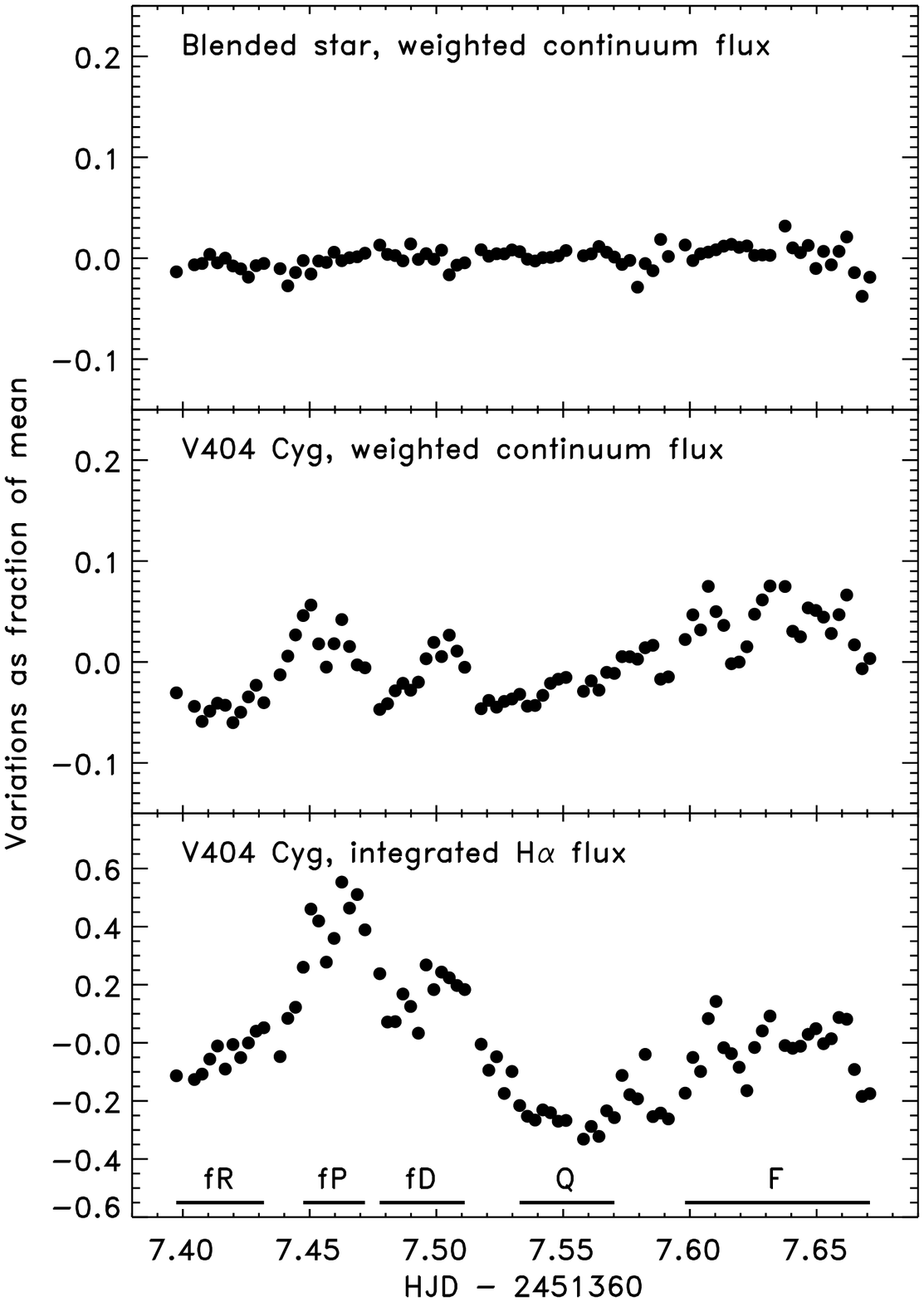}{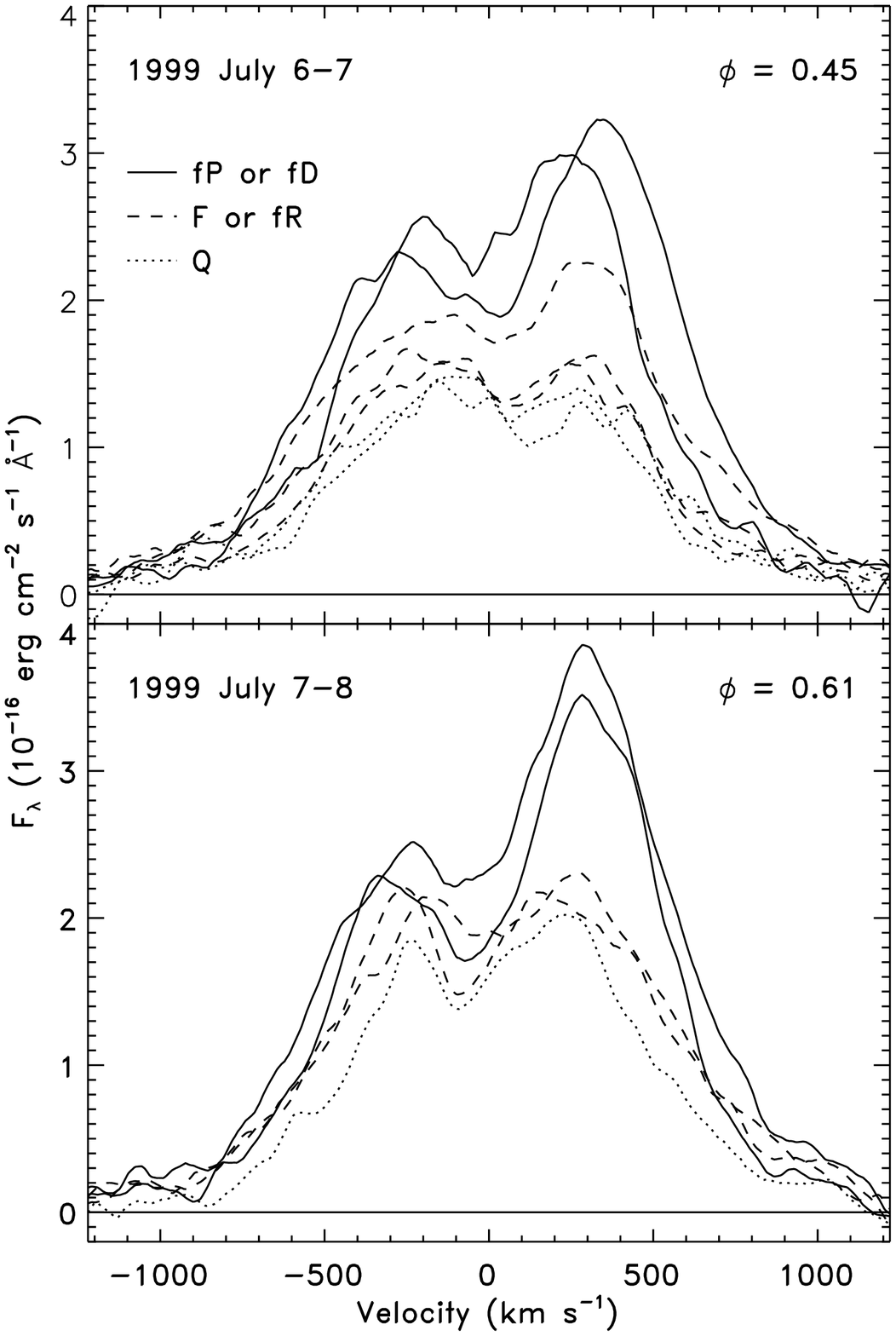}
\caption{Left: Line and continuum lightcurves from 1999 July 7-8.
Right: Line profiles for different states from both nights.  The
abbreviations are Q = quiescent; faintest state, F = flickering state,
fR, fP and fD = flare rise, peak and decline respectively.}
\end{figure}

We find significant line and continuum variability; the H$\alpha$ line
is particularly dramatic showing variations of a factor of two in flux
over a few hours, much less than the 6.5\,day orbital period.  The
line and continuum variations generally appear well correlated with no
detectable lag between them ($<4$\,min).  

During the large flares, significant profile changes are seen.  A
response is seen across the whole line profile and the excess light
has a double peaked profile, suggesting that the whole disc is
involved.  Interestingly the excess light is not symmetric and the red
wing is usually significantly stronger (in archival data as well as
that presented here), independent of orbital phase.  This suggests
that it does not indicate the kinematics of a brighter region but that
some other effect, perhaps vertical motion or absorption of the blue
wing by approaching material, is at play.

It seems likely that the H$\alpha$ emission is driven by X-ray flares;
hence the whole disc can be affected on short timescales.  A wind
driven by the flares could then be responsible for absorbing the blue
wing of the profile.

\bigskip
\acknowledgements{RIH, CAH and PAC acknowledge
financial support the Leverhulme Trust.  The WHT and
JKT are operated on La Palma by the ING in the Spanish Observatorio
del Roque de los Muchachos of the IAC.}


\begin{references}
\reference
Casares J., Charles P. A., 1992, MNRAS, 255, 7
\reference
Casares J., Charles P. A., Naylor T., Pavlenko E. P., 1993, 
MNRAS, 265, 834	
\reference
Hynes R. I., Zurita C., Haswell C. A., Casares J., Charles P.,
Pavlenko E. P., Shugarov S. Yu., Lott D., 2002, MNRAS, submitted
\end{references}
\end{document}